\documentclass[aps,superscriptaddress,showpacs,nofootinbib,eqsecnum]{revtex4}

\usepackage{epsfig}  \input{epsf}
\def\be{\begin{equation}} \def\ee{\end{equation}} \def\bea{\begin{eqnarray}}
\def\eea{\end{eqnarray}}

\def\beq{\begin{equation}}
\def\eeq{\end{equation}}
\def\beqa{\begin{eqnarray}}
\def\eeqa{\end{eqnarray}}

\begin{document}

\title{Chiral Transition Within Effective Quark Models Under Magnetic Fields}

\author{Gabriel N. Ferrari} 
\affiliation{Departamento de F\'{\i}sica, Universidade Federal de Santa
  Catarina, 88040-900 Florian\'{o}polis, Santa Catarina, Brazil}

\author{Andre F. Garcia} 
\affiliation{Departamento de F\'{\i}sica, Universidade Federal de Santa
  Catarina, 88040-900 Florian\'{o}polis, Santa Catarina, Brazil}

\author{Marcus B. Pinto} \email{marcus@fsc.ufsc.br}
\affiliation{Departamento de F\'{\i}sica, Universidade Federal de Santa
  Catarina, 88040-900 Florian\'{o}polis, Santa Catarina, Brazil}

\begin{abstract}
We consider the simplest versions of  the  Nambu--Jona-Lasinio (NJL) model and the Linear Sigma Model (LSM), in the Mean Field Approximation (MFA),  in order to analyze hot and dense two flavor quark matter subject to strong magnetic fields. We pay especial attention to the case of a finite chemical potential,  which has not yet been fully explored.  Our results, for the NJL model, are in qualitative  agreement with other  recent applications  showing that, for stronger fields, the first order segment of the  transition line increases with the magnetic strength  while the coexistence chemical potential value, at low temperatures, decreases. In the present work, one of the most important results is related to the analysis of how these features affect the phase coexistence region in the  $T-\rho_B$ plane. We find that the coexistence boundary oscillates around the $B=0$ value for magnetic fields of the order $eB \lesssim 9.5\, m_\pi^2$ which can be understood by investigating the filling of Landau levels at vanishing temperature. So far, most investigations have been concerned with the effects of the magnetic field over the $T-\mu$ plane only  while other thermodynamical quantities such as the adiabats, the quark number susceptibility, the interaction measure and the latent heat have been neglected. Here, we take a step towards filling this gap by investigating  the influence of a magnetic field over these quantities. Finally, we argue that a naive application of the MFA does not seem to be appropriate to treat the LSM in the presence of magnetic fields.  
\end{abstract}

\pacs{11.10.Wx, 26.60.Kp,21.65.Qr, 25.75.Nq, 12.39.Ki}

\maketitle

\section{Introduction}

The determination of the QCD phase diagram, even in the absence of magnetic fields, is still a
matter of great theoretical and experimental activities. In this case,
powerful lattice simulations have established that at vanishing baryon
chemical potential there is no true phase transition from hadronic matter to a quark gluon-plasma but rather a very rapid raise in the energy density signaling a crossover characterized by a pseudocritical temperature, $T_{\rm pc}\approx160\,{\rm MeV}$ \cite{Aoki}. The situation is less clear for the finite chemical potential region since, so far, there is no reliable information avaliable from lattice QCD  evaluations. Nevertheless, most finite $\mu$ lattice extrapolations for the $\mu=0$ Columbia plot indicate that the critical first order surface (in the $m_{u,d}-m_s-\mu$ space) will hit the physical current mass values at some finite $\mu$ thereby characterizing a critical end point (CP) \cite {fukuhatsuda} which constitutes the most plausible theoretical scenario. On the other hand, other lattice evaluations \cite {deForcrand} predict that, as $\mu$ increases, this critical first order surface bends in a way which in principle would exclude the appearance of a CP and the whole $T-\mu$ plane would then be dominated by a crossover  representing an ``exotic'' scenario \cite {fukuhatsuda}. However, even this  situation may be reversed, in favor of a critical end point, if a different physics strongly influences the finite density region causing a back bending of the critical surface \cite {Fukushima,kapusta,fkp}.

At the same time, the possibility that  strong magnetic fields may be produced  in non central heavy ion collisions \cite {kharzeev09}  as well as being present in magnetars \cite {magnetars}, and in the early universe \cite {universe} leads to the question of how these fields influence the QCD phase diagram. So far, most estimates have been carried out at vanishing chemical potential  with the aid of effective theories such as the Linear Sigma Model (LSM) \cite {eduardo} and the Nambu--Jona-Lasinio model (NJL) \cite {ruggieri,hotnjl} within the Mean Field Approximation (MFA). The general outcome is that the pseudocritical temperature, at which the crossover takes place, increases with increasing values of the magnetic field. Some of these  applications have also considered these effective models with the inclusion of the Polyakov loop to take confinement into account. The results show that  the $T_{\rm pc}$ related to deconfinement also increases with $B$ but an interesting splitting between this temperature and the one related to chiral symmetry has been observed \cite{eduardo}. At $\mu=0$, the first lattice attempt to solve the problem considered two quark flavors and high values of pion masses ($m_\pi=200-400 \, {\rm MeV}$)  confirming that    $T_{\rm pc}$ should increase with $B$ \cite {earlylattice}. 
However, an improved lattice simulation \cite {lattice} which considered 2+1 quark flavors at physical pion mass values ($m_\pi=140 \, {\rm MeV}$), together with an  extrapolation to the continuum, predicted that  $T_{\rm pc}$ should  decrease with $B$. A decrease of $T_{\rm pc}$, for the deconfinement transition, has also been observed with the MIT bag model \cite {mit} and, more recently, with the large-$N_c$ limit of QCD \cite {neweduardo} (see Ref. \cite {edureview} for a summary of these results) .  In Ref. \cite{fukushima2012} it is
suggested that this behavior is due to the entanglement with the
Polyakov loop that is affected by magnetic screening as discussed in
Ref. \cite{galilo}, where it was shown,  within a simplified framework, that the strong electromagnetic fields
can play a catalysing role for a deconfinement transition. However, as
referred in Ref. \cite{fukushima2012}, if only the
chiral sector is considered, an
enhancement of the chiral condensate, due to finite $B$, would not be contradictory
with the lattice simulations of Ref. \cite{lattice}.
Nevertheless, although the quantitative discrepancy remains to be fully understood,  there is a concensus regarding the crossover character of the chiral transition at vanishing chemical potential.  On the other hand, the effects of strong magnetic fields at the  finite $\mu$ regime has not been fully explored to date even within effective theories. 
Experimentally, the intermediate temperature regime is important regarding current nuclear collisions at the low-energy end of RHIC 
and in the future with FAIR at GSI and NICA at JINR whose aim is to probe the expected critical point region. At the low temperature end  one expects to find a rather interesting physics related to the expected first order phase transition which may have astrophysical consequences (e.g., regarding the possibility of quark star formation \cite {prcsu2,prcsu3}).

In the chiral limit, the two flavor version of the NJL model subject to a magnetic field was  considered in Ref. \cite {inagaki}  with the MFA while the three flavor version, at the physical point, was recently analyzed in Ref. \cite {nois} within the same approximation. In the latter reference, which concerns the more realistic case, it was observed that $T_{\rm pc}$, at $\mu=0$, always increases with $B$, as in most model applications.  In that work, one of  the  main novelties regards the critical end point location, which moves towards higher temperatures and smaller chemical potential values as $B$ increases. Another interesting result concerns  the size of the first order segment of the transition line, which becomes longer, as stronger magnetic fields are considered. Also, at low temperatures, it has been observed that    the  chemical potential value at which the first order transition occurs decreases with $B$. The latter result has been previously observed with the two flavor NJL, in the chiral limit  \cite {inagaki}, as well as with  a holographic one-flavor model \cite {andreas} .  The interested reader may find a model-independent physical explanation for this result, which the authors have termed Inverse Magnetic Catalysis (IMC), in Ref. \cite {andreas} while a recent review with new analytical results for the NJL can be found in Ref. \cite {imc}. Also, very recently, the two flavor NJL model has been considered to investigate the dynamics of neutron mesons in a hot and magnetized medium \cite{mesons}. Another important application regards the generalized three flavor NJL model including eight-quark interactions in which the phenomenon of secondary magnetic catalysis sets in \cite {brigitte}.

The LSM with two flavors  has also been employed to determine the $T-\mu$ phase diagram with the MFA \cite {andkhan} and, more recently, with the more powerful Functional Renormalization Group (FRG) \cite{newandersen}. This method has been previously used  in an application to the Polyakov quark model (PQM) demonstrating that, at $\mu=0$, the increase of $T_{\rm pc}$ with $B$ persists even when mesonic fluctuations are considered \cite {skokov}. The MFA results of Ref. \cite {andkhan} show that, when all fermionic contributions are considered, the whole $T-\mu$ plane is dominated by the crossover, irrespective of the magnetic field value while the chiral symmetry broken region expands with $B$. 
It is interesting to note here that, in this case, the LSM mimics the ``exotic'' theoretical QCD phase diagram scenario mentioned at the beginning of this section. However, as we show here, this is just an artifact of the parametrization adopted in Ref. \cite {andkhan} where the authors chose a rather high value for the sigma meson mass ($m_\sigma=800 \, {\rm MeV}$). As a matter of fact,  the occurrence of a critical point and first order transition depends crucially on the mass of the sigma meson that one uses in the computation \cite {sandeep}. Apart from the parametrization issue one must be very careful when considering the LSM since different approximations may lead to very different results even at the qualitative level (e.g., yielding different types of phase transitions \cite{andkhan,friman,akk}). This is mainly due to the fact that within this effective theory one has scalar and pseudoscalar mesons as well as quarks as degrees of freedom. Also, the pure vacuum contributions may be neglected without spoiling the breaking of chiral symmetry, which happens at the classical level. This allows for different approaches in which one or more contributions are neglected yielding different results. 
When magnetic fields are present, the phase diagram obtained with the FRG applied to the LSM  \cite{newandersen} agrees, qualitatively,  with the ones obtained with the NJL in the MFA  \cite {nois, mesons}. However, sigma masses of the order $m_\sigma= 400-450 \, {\rm MeV}$ have been considered so that one may wonder  if the qualitative agreement  between the LSM-FRG  and the NJL-MFA predictions for the chiral transition phase diagram are due to the use of smaller $m_\sigma$ values or to the use of a more powerful approximation scheme within the LSM. Here, in order to address this question, we consider  the LSM with the MFA using $m_\sigma=600\, {\rm MeV}$ and also $m_\sigma=450\, {\rm MeV}$ showing that the higher value reproduces the ``exotic'' scenario also found in Ref. \cite {andkhan}. On the other hand, using $m_\sigma=450\, {\rm MeV}$, we observe the appearance of a first order line starting at $T=0$ and terminating at the critical end point. However, when magnetic fields are turned on, we do not observe the same qualitative features observed with the NJL-MFA and the LSM-FRG at intermediate to low values of the temperature. For instance, at $T=0$, the coexistence chemical potential values does not decrease with $B$ as predicted by the Inverse Magnetic Catalysis mechanism \cite {andreas,imc} which points out to the fact that the naive MFA application as performed here, and also in Ref. \cite {andkhan}, may not be adequate to treat the LSM in the presence of magnetic fields.

Finally, note that despite all the progress made so far in analyzing the influence of $B$ over the $T-\mu$ phase diagram not much effort has been devoted to investigate particular aspects such as the coexistence and spinodal boundaries associated with the first order transition. Also, the analysis of quantities such as the quark number susceptibility, adiabats, interaction measure, latent heat, etc  may help in the understanding of the phase diagram structure under strong magnetic fields. Therefore, the main goal of the present work is to improve over the simple $T-\mu$ phase diagram by investigating its structure and related physical quantities in more detail. For instance, to the best of our knowledge, the  influence of $B$ over the  coexistence  and the critical regions, for example, has not been addressed before. 

In this work we consider the  LSM and the NJL model with two flavors in the framework of the MFA  in an application which could be viewed as an extension of Ref \cite {scavenius} to  the $B \ne 0$ case (the thermdoynamics of both models, at $B=0$, has been recently discussed with great detail in Ref. \cite {tiwari}). Note that, here, our main focus is on the NJL which yields the expected phase diagram scenario, even at the MFA level, allowing us to probe the physically rich first order transition region with its  associated critical end point. On the other hand, as already mentioned, the MFA treatment of the LSM at $B \ne 0$ does not generate the expected phase diagram in the full $T-\mu$ plane by excluding the IMC phenomenon for example. Here, its investigation is justified 
for  allowing us to  trace this problem as being generated by the MFA and not only by a parametrization which considers high $m_\sigma$ values. These two models, extended by the Polyakov loop, have recently been considered  in Ref. \cite {newmarco} where quantities like the magnetic susceptibility of the quark condensate as well as the quark polarization have been evaluate at vanishing temperature.

In the next section we investigate the phase structure of the NJL paying special attention to the low temperature regime which has been less explored in the literature. In the same section we evaluate the effect of  $B$ over the  coexistence and spinodal regions,  isentropic lines and the quark susceptibility with the aid of the $T-\mu$, $T-\rho_B$ and $P-T$  planes. In Section 3, we consider the LSM in the MFA framework, as  in  Ref. \cite {andkhan}, but extending the analysis  so as to consider the effects of $B$ in the sigma meson mass. We also present the $T-\mu$ phase diagram for this model, at the physical point, for the magnetic field values relevant to RHIC and the LHC. This exercise will allow us to compare the MFA results furnished by both models. Our conclusions are presented in Section 4.

\section{Thermodynamics of the  NJL under a Magnetic Field}

At vanishing temperature and finite density the NJL model, subject to a magnetic field, has been used  to address different questions such as the stability of quark droplets  \cite{klimenko} and  the EoS of magnetars with  and without strangeness in Refs. \cite {prcsu3} and \cite {prcsu2} respectively. The influence of $B$ and instantons at the two extreme parts of the phase diagram ($T=0, \mu \ne0$ and $T \ne 0, \mu=0$) was studied in Ref. \cite {BB}. The whole $T-\mu$ plane was first analyzed in Ref. \cite {inagaki} where only the chiral limit, for the two flavor version, was considered.  At the physical point,  the same version of the model has been recently considered in Ref. \cite {mesons} whose results, for the phase diagram, qualitatively agree with the ones found in Ref. \cite {nois} for three flavors. In this Section we will  obtain the EoS and then analyze different physical quantities in order to  characterize the physical situations  which are more sensitive to magnetic effects.

\subsection{The Model}

The NJL model is described by a Lagrangian density for fermionic fields given by~\cite{njl}

\begin{equation}
\mathcal{L}_{\rm NJL}={\bar \psi}\left( i{\partial \hbox{$\!\!\!/$}}-m\right) \psi
+G\left[ ({\bar \psi}\psi)^{2}-({\bar{\psi}} \gamma _{5}{\vec{\tau}}\psi
  )^{2}\right] ,
\label{njl2}
\end{equation}
\noindent
where $\psi$ (a sum over flavors and color degrees of freedom is implicit)
represents a flavor iso-doublet ($u,d$ type of quarks) $N_{c}$-plet quark
fields, while $\vec{\tau}$ are isospin Pauli matrices. The Lagrangian density
(\ref{njl2}) is invariant under (global) $U(2)_{\rm f}\times SU(N_{c})$ and,
when $m=0$, the theory is also invariant under chiral $U(2)_{L}\times
U(2)_{R}$. 

Due to the quadratic fermionic interaction, the theory is nonrenormalizable
in 3+1 dimensions ($G$ has dimensions of $\mathrm{eV}^{-2}$), meaning that
divergences appearing at successive perturbative orders cannot be all
eliminated  by a consistent redefinition of the original model parameters
(fields, masses, and couplings). The renormalizability issue arises during the
evaluation of momentum integrals associated with loop Feynman graphs in a
perturbative expansion and, in the process, one usually employs regularization
prescriptions (e.g. dimensional regularization, sharp cut-off, etc) to
formally isolate divergences. However, the procedure introduces
\textit{arbitrary} parameters with dimensions of energy that do not appear in
the \textit{original} Lagrangian density.
Within the NJL model a sharp cut off ($\Lambda$) is preferred and since the
model is nonrenormalizable, one has to  fix $\Lambda$ to a value related to the
physical spectrum under investigation. This strategy turns the 3+1 NJL model
into an effective model, where $\Lambda$ is treated as a parameter. The phenomenological values
of quantities such as the pion mass ($m_{\pi}$),  the pion decay constant
$(f_{\pi})$, and the quark condensate ($ \langle {\bar \psi} \psi \rangle$) are used to fix  $G$, $\Lambda$, and 
$m$.  Here, we adopt the values $m= 6 \, {\rm MeV}$, $\Lambda= 590 \, {\rm MeV}$ and $G \Lambda^2= 2.435$ which have also
been employed in Ref. \cite {BB} (see Refs. \cite {buballa,opt} for other possibilities).

\subsection {The NJL free energy in the presence of a magnetic field}

The free energy,  in the MFA, can be written as follows \cite {buballa,prcsu2} (see Ref. \cite {opt} for results beyond MFA)

\begin{equation} \label{Veff_NJL}
\mathcal{F}^{\rm NJL}=\frac{(M-m)^2}{4G}+\frac{i}{2}{\rm tr} \int \frac{d^4p}{(2\pi)^4} \ln  [-p^2+M^2] \,\,,
\label{free}
\vspace{0.4 cm}
\end{equation}

\noindent where $M$ is the constituent quarks mass. In order to study the effect of a magnetic field in the chiral transition at finite temperature and chemical potential a dimensional reduction is induced  via  the following replacements  in Eq. (\ref{Veff_NJL}):

\begin{equation}
p_0\rightarrow i(\omega_{\nu}-i\mu)\,\, , \nonumber
\end{equation}

\begin{equation}
p^2 \rightarrow p_z^2+(2n+1-s)|q_f|B \,\,\,\,\, , \,\,\mbox{with} \,\,\, s=\pm 1 \,\,\, , \,\, n=0,1,2...\,\,, \nonumber
\end{equation}

\begin{equation}
\int_{-\infty}^{+\infty}  \frac{d^4p}{(2\pi)^4}\rightarrow i\frac{T |q_f| B}{2\pi}\sum_{\nu=-\infty}^{\infty}\sum_{n=0}^{\infty}\int_{-\infty}^{+\infty} \frac{dp_z}{2\pi} \,\,,\nonumber
\end{equation}

\noindent where $\omega_\nu=(2\nu+1)\pi T$, with $\nu=0,\pm1,\pm2...$ represents the Matsubara frequencies for fermions, $n$ represents the Landau levels and $|q_f|$ is the absolute value of the quark electric charge ($|q_u|= 2e/3$, $|q_d| = e/3$ with $e = 1/\sqrt{137}$ representing the electron charge).  Following Ref. \cite {prcsu2} we can write the  free energy as

\begin{equation}
\mathcal{F}^{\rm NJL}=\frac{(M-m)^2}{4G} + \mathcal{F}_{\rm vac}^{\rm NJL}+\mathcal{F}_{\rm mag}^{\rm NJL}+\mathcal{F}_{\rm med}^{\rm NJL} \,\,\,\,,
\end{equation}

\noindent where

\begin{equation} \label{Vvac}
\mathcal{F}_{\rm vac}^{\rm NJL}=-2N_cN_f \int \frac{d^3{\bf p}}{(2\pi)^3}({\bf p}^2 +M^2)^{1/2} \,\,\,.
\label{vacnjl}
\end{equation}
This divergent integral is regularized by a sharp cut-off, $\Lambda$, yielding
\begin{equation}
\mathcal{F}_{\rm vac}^{\rm NJL}= \frac{N_c N_f}{8\pi^2} \left \{ M^4 \ln \left [
    \frac{(\Lambda+ \epsilon_\Lambda)}{M} \right ]
 - \epsilon_\Lambda \, \Lambda\left[\Lambda^2 +  \epsilon_\Lambda^2 \right ] \right \}\,\,,
\end{equation}
where we have defined $\epsilon_\Lambda=\sqrt{\Lambda^2 + M^2}$. 
The magnetic and the in-medium terms are respectively given by 
\begin{equation} \label{Vmag}
\mathcal{F}_{\rm mag}^{\rm NJL}=-\frac{N_c}{2\pi^2}\sum_{f=u}^d (|q_f|B)^2\left\{\zeta^{(1,0)}(-1,x_f)-\frac{1}{2}[x_f^2-x_f]\ln(x_f)+\frac{x_f^2}{4} \right\}\,\,\,,
\end{equation}
and
\begin{equation} \label{Vmed}
\mathcal{F}_{\rm med}^{\rm NJL}=-\frac{N_c}{2\pi}\sum_{f=u}^d\sum_{k=0}^{\infty}\alpha_k |q_f|B \int_{-\infty}^{+\infty} \frac{dp_z}{2\pi}\left\{T \ln [1+{e}^{-[E_{p,\,k}(B)+\mu]/T}]+ T \ln [1+{e}^{-[E_{p,\,k}(B)-\mu]/T}]\right\} \,.
\end{equation}

\noindent Note that in the last equation we have replaced the label $n$ by $k$ in the Landau levels in order to account for the degeneracy factor $\alpha_k=2-\delta_{0k}$. In Eq. (\ref{Vvac}), $N_c=3$ and $N_f=2$ are the color and flavor degrees of freedom, respectively. Also,  in Eq (\ref{Vmag}) we have used $x_f=M^2/(2|q_f|B)$ and $\zeta^{(1,0)}(-1,x_f)=d\zeta(z,\,x_f)/dz|_{z=-1}$ with $\zeta(z,\,x_f)$ representing  the Riemann-Hurwitz function (the details of  the manipulations leading to the equations above can be found in the appendix of Ref. \cite {prcsu2}). Finally, in Eq. (\ref{Vmed}) we have $E_{p,\,k}(B)=\sqrt{p_z^2+2k|q_f|B +M^2}$ where  $M$  is  the effective
self consistent quark mass 

\begin{eqnarray}
M &=& m +\frac{ N_c N_f MG}{\pi^2} \left \{
\Lambda
\sqrt {\Lambda^2 + M^2} -\frac{M^2}{2}
\ln \left [ \frac{(\Lambda+ \sqrt {\Lambda^2 + {M^2}})^2}{{M}^2} \right ] \right \} \nonumber \\
&+&\frac{N_c MG}{\pi^2}\sum_{f=u}^d  |q_f| B\left \{ \ln [\Gamma(x_f)] 
-\frac {1}{2} \ln (2\pi) +x_f -\frac{1}{2} \left ( 2 x_f-1 \right )\ln(x_f) \right \} \nonumber \\
&-& \frac{N_cMG}{2\pi^2}\sum_{f=u}^d \sum_{k=0}^{\infty} \alpha_k |q_f|B \int_{-\infty}^{\infty} \frac{dp_z}{E_{p,k}(B)}\left\{ \frac{1}{e^{[E_{p,k}(B)+\mu]/T}+1}+\frac{1}{e^{[E_{p,k}(B)-\mu]/T}+1} \right\} \,\,.
\label{MmuB}
\end{eqnarray}
\begin{figure}[tbh]
\vspace{0.5cm} 
\epsfig{figure=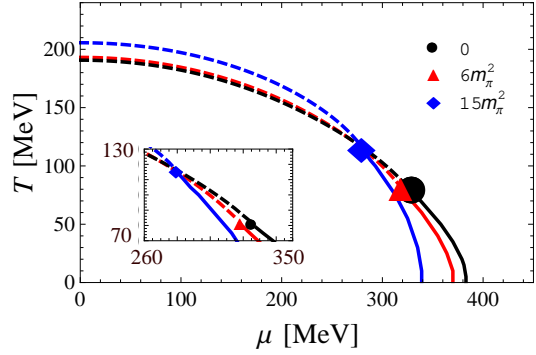,angle=0,width=7cm}
\caption{ Phase diagram for the NJL model in the $T-\mu$ plane. The CP, represented by solid symbols, occurs at ($T_c\simeq 78.5 \, {\rm MeV}, \mu_c\simeq 328\, {\rm MeV}$) for $B=0$, ($T_c\simeq 81.7 \, {\rm MeV}, \mu_c\simeq 318\, {\rm MeV}$) for  $eB=6 \, m_\pi^2$, and ($T_c\simeq 115.6 \, {\rm MeV}, \mu_c \simeq 279.3\, {\rm MeV}$) for  $eB=15\, m_\pi^2$. The continuous lines represent  first order phase transitions and the dashed lines represent crossovers.  The inset shows the crossing of the transition lines.}
\label{fig1}
\end{figure}

\begin{figure}[tbh]
\vspace{0.5cm} 
\epsfig{figure=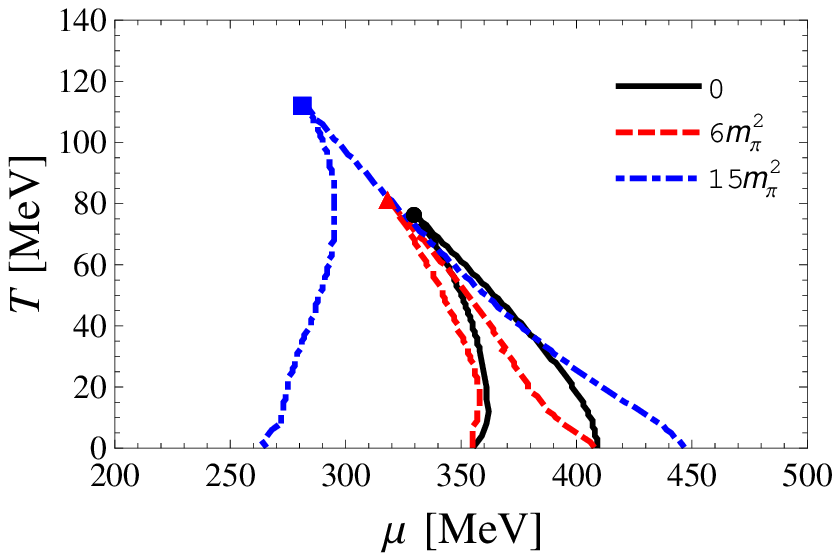,angle=0,width=7cm}
\epsfig{figure=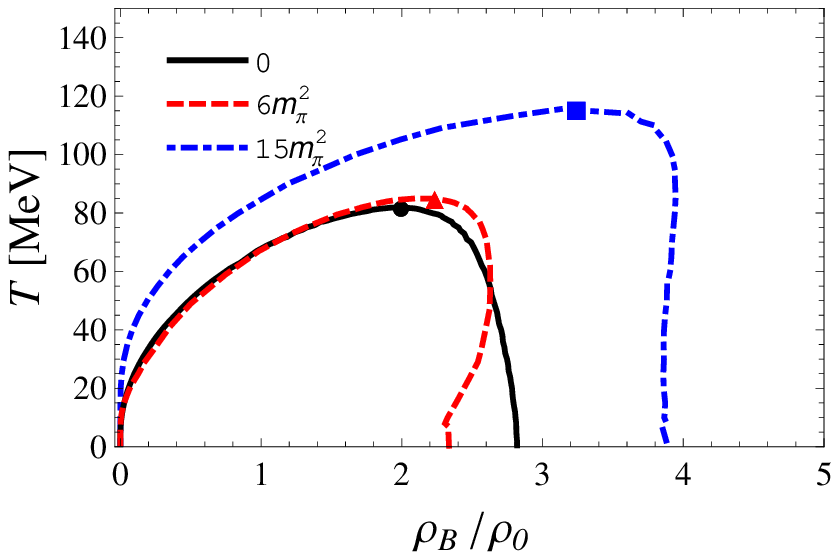,angle=0,width=7cm}
\caption{ Left panel: Spinodal boundaries for the NJL model  in the $T-\mu$ plane. Right panel: Phase coexistence boundaries in the $T-\rho_B$ plane ($\rho_B$ appears in units of the nuclear matter density, $\rho_0=0.17 \, {\rm fm}^{-3}$). The solid symbols indicate the location of the critical point for each value of $B$. }
\label{fig2}
\end{figure}
\subsection{Numerical Results for the NJL model} 

We can now obtain the phase diagram as well as other important physical quantities for the NJL model in the presence of $B$. All the relevant thermodynamical quantities can be readily obtained by recalling that the free energy, evaluated at the mass value which satisfies the gap equation, gives the negative of the pressure, ${\cal F}(M)=-P$. Then, the net quark number density  is obtained from $\rho= dP/d\mu$, and the entropy density from $s= dP/dT$ while the energy density is ${\cal E}= -P + T s + \mu \rho$. 
Let us start with the phase diagram, in the $T-\mu$, as shown by Fig. \ref{fig1} which was obtained for vanishing $B$ as well as for $eB= 6 m_\pi^2$ and $eB= 15 m_\pi^2,$ which cover the estimated values for non central collisions at RHIC and the LHC, respectively \cite{estimates}. At $\mu=0$ one observes an increase of $T_{\rm pc}$ which was expected to happen within the present model approximation, as discussed in the introduction. However, at around $\mu \approx 285 \, {\rm MeV}$  the $T_{\rm pc}$ value for $eB=6\, m_\pi^2$  is smaller than the $B=0$ value,  in accordance with the findings of Refs. \cite {nois,mesons, newandersen}. The figure also indicates that $B$ induces a  noticeable increase of the first order segment of the transition line which terminates at higher values of $T$ and smaller values of $\mu$. At low temperatures,  the coexistence values of the chemical potential decrease with increasing $B$ showing IMC (although $\mu$ starts to increase again for values $eB \gtrsim 16 m_\pi^2 \approx \Lambda^2$ \cite{nois,mesons} which we do not consider here). At $T=0$ this pattern was also observed in Ref. \cite {klimenko}. The CP is located at  ($T_c \simeq 78.5 \, {\rm MeV}, \mu_c\simeq 328\, {\rm MeV}$) for $B=0$, ($T_c \simeq 81.7 \, {\rm MeV}, \mu_c \simeq 318\, {\rm MeV}$) for $eB=6 \, m_\pi^2$, and ($T_c \simeq 115.6 \, {\rm MeV}, \mu_c \simeq 279.3\, {\rm MeV}$) for $eB=15\, m_\pi^2$.

\begin{figure}[tbh]
\vspace{0.5cm} 
\epsfig{figure=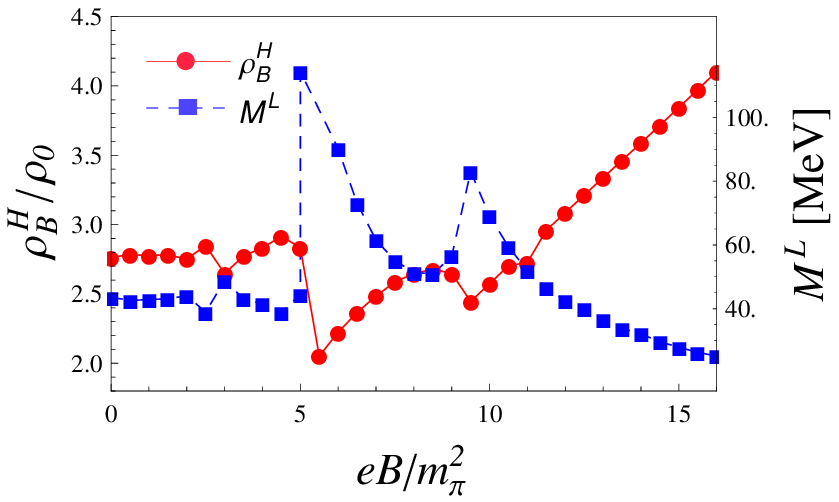,angle=0,width=8cm}
\epsfig{figure=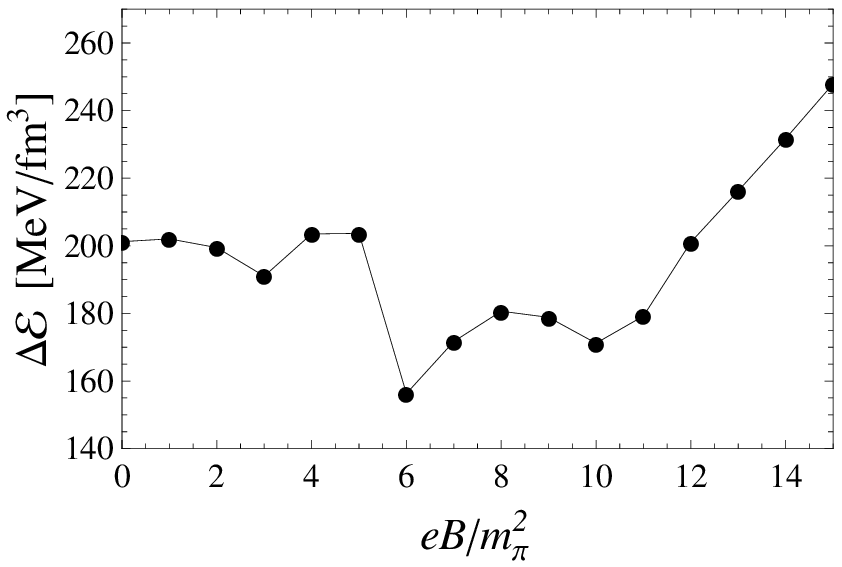,angle=0,width=7cm}
\caption{ Left panel: The NJL model effective quark mass (squares) at the lowest value occurring at the transition, $M^L$, and the highest coexisting baryon density (dots), $\rho^H_B$ (in units of $\rho_0$), as functions of $eB/m_\pi^2$ at $T=0$. The lines are shown  just in order to guide the eye. Right panel: The latent heat, $\Delta {\cal E}$, at $T=0$, as a function of $eB/m_\pi^2$. As expected, this quantity and $\rho^H_B$ behave in  a similar way. }
\label{fig3}
\end{figure}
\begin{figure}[tbh]
\vspace{0.5cm} 
\epsfig{figure=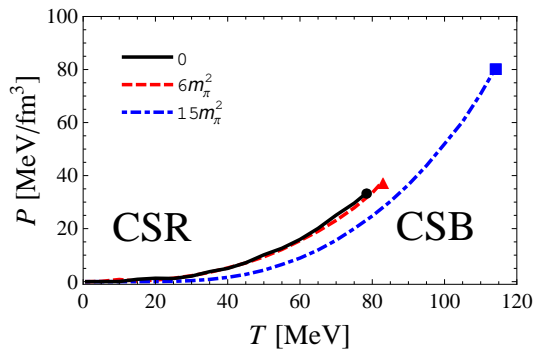,angle=0,width=7cm}
\caption{ The NJL model phase diagram in the  $P-T$ plane indicating the regions of broken, CSB, and (partially) restored chiral symmetry, CSR, which correspond to the ``gas'' and ``liquid'' phase respectively. All lines represent first order phase transitions which terminate at the CP represented by the solid symbols.}
\label{fig4}
\end{figure}

Fig 2 (left panel) displays the spinodal lines, also in the $T-\mu$ plane, for the same values of the magnetic field  showing that very high fields increase significantly the size of the metastable regions.  
In the $T-\rho_B$ plane, the coexistence boundary is limited by the two distinct, low ($\rho^L$) and high ($\rho^H$), densities with $\rho^L < \rho^H$ for $T<T_c$ and $\rho^L = \rho^H$ at $T=T_c$. The boundaries, for different values of $B$, are shown in Fig 2 (right panel) where we work in terms of the baryon density, $\rho_B=\rho/3$, given in units of the nuclear matter density, $\rho_0 = 0.17 \, {\rm fm}^{-3}$. An interesting  feature of this  figure is the oscillation of the coexistence boundary around its value at $B=0$, which is more pronounced in the $\rho^H$ branch.  The decrease in $\rho^H$ for $eB=6m_{\pi}^2$, at low temperature, shown in the right panel of Fig 2 can be understood in terms of the filling of the Landau levels and, with this aim, we present Fig 3 (left panel) which displays the baryonic density and the effective quark mass  as  functions of the magnetic field at $T=0$. To analyze the figure let us recall that, in the limit $T\rightarrow 0$, the baryonic density can be written\footnote {There is a misprint in Eq. (30) of Ref. \cite {prcsu2} where it should be $\rho_B$ instead of $\rho$.} as \cite {prcsu2}

\begin{equation}
 \rho_B (\mu,B)= \theta (k_F^2)\sum_{f=u}^d \sum_{k=0}^{k_{f,max}} \alpha_k \frac{|q_f| B N_c}{6\pi^2}k_F \,\,,
\label{eq_rho_t0}
\end{equation}
where $k_F=\sqrt{\mu^2-2|q_f|kB-M^2}$ and
\begin{equation} 
k_{f,max} = \frac{\mu^2 - M^2}{2|q_f|B} \,,
\end{equation}
or the nearest integer.
\begin{figure}[tbh]
\vspace{0.5cm} 
\epsfig{figure=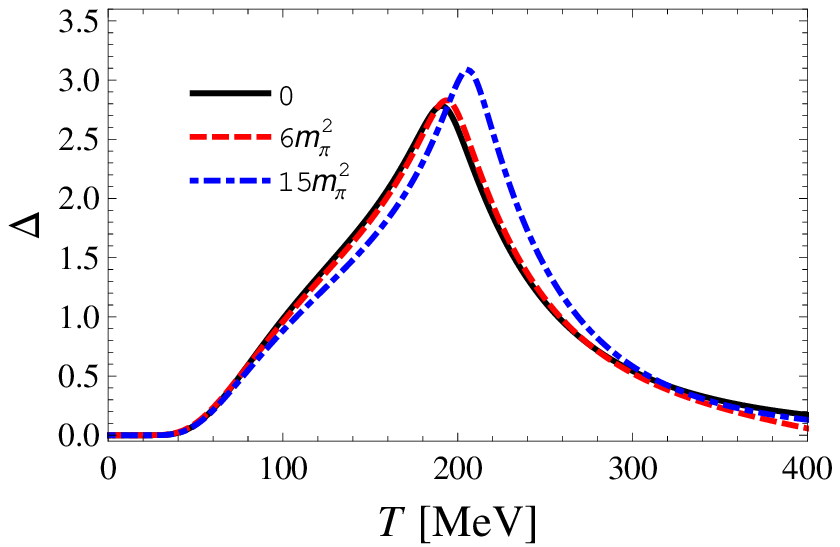,angle=0,width=7cm}
\epsfig{figure=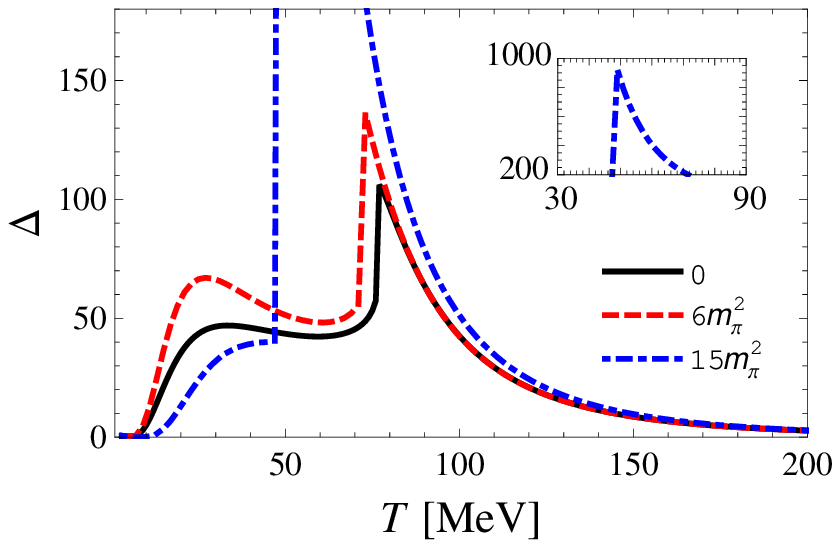,angle=0,width=7cm}
\caption{Left panel: Interaction measure for the NJL model as a function of $T$, for $\mu=0$, displaying the crossover behavior. Right panel: Interaction measure for the NJL model as a function of $T$, $\mu=\mu_c(B=0)= 328 \, {\rm MeV}$ displaying the first order transition behavior.}
\label{fig5}
\end{figure}

\begin{figure}[tbh]
\vspace{0.5cm} 
\epsfig{figure=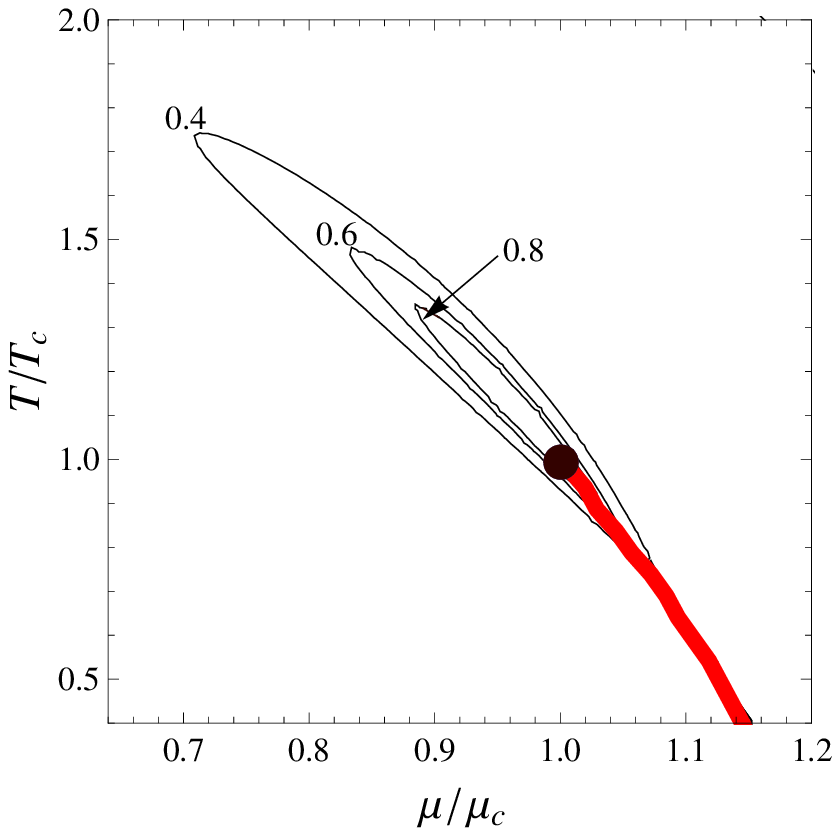,angle=0,width=6cm}
\epsfig{figure=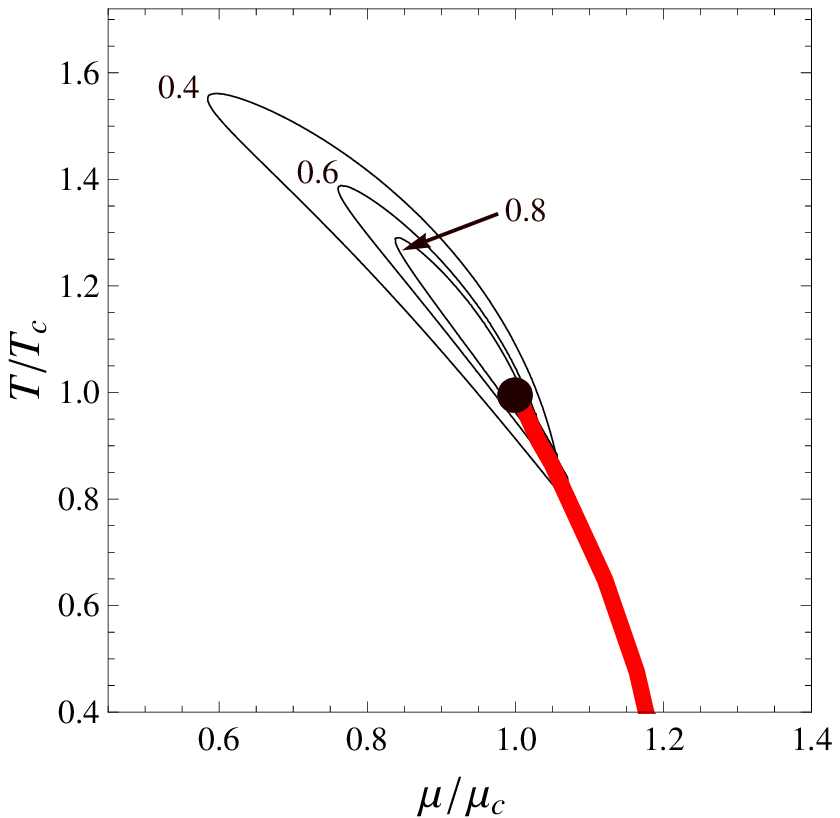,angle=0,width=6cm}
\caption{Critical regions for the NJL model at $eB=0$ (left panel) and $eB=15\,m_\pi^2$ (right panel) given by the contour lines of the quark number susceptibility, $\chi_q$, in units of $\Lambda^2$. The control parameters have been normalized by $T_c$ and $\mu_c$ corresponding to each value of the magnetic field. In both cases the critical region is elongated in the direction parallel to the first order transition line.}
\label{fig6}
\end{figure}
Eq. (\ref {eq_rho_t0}) shows that if $k_F^2 <0$ then $\rho_B=0$ which is precisely the low density value at $T=0$ which is easy to understand by recalling that the effective mass is double valued when the first order transition occurs presenting a high ($M^H$) and a low ($M^L$) value with  $M^L < M^H$ for $T<T_c$ and $M^L = M^H$ at $T=T_c$. Now, at $T=0$, $M^H$ corresponds to the value effective quark mass acquires when $T=0$ and $\mu=0$ (the vacuum mass) which corresponds to $M^H \simeq 403\, {\rm MeV}$ at $B=0$, $M^H \simeq 416\, {\rm MeV}$ at $eB=6 \, m_\pi^2$, and $M^H \simeq 467\, {\rm MeV}$ at $eB=15 \, m_\pi^2$. On the other hand, at $T=0$ the first order transition happens when  $\mu \simeq 383\, {\rm MeV}$ for $B=0$, $\mu \simeq 370 \, {\rm MeV}$ for $eB=6 \, m_\pi^2$ and 
$\mu \simeq 339 \, {\rm MeV}$ for $eB=15 \, m_\pi^2$ so that $\rho^L=0$ even at the lowest Landau level, as required by $\theta(k_F^2)$ in Eq. (\ref {eq_rho_t0}). Then, to understand the oscillations let us concentrate on the $\rho^H$ branch which is shown, together with $M^L$ (the in-medium mass), in Fig. 3 (left panel) where it is clear that both quantities have an opposite oscillatory behavior. The origin of the oscillations in these quantities can be traced back to the fact that  $k_{max}$ (the upper Landau level filled) decreases as the magnetic field increases. The first and second peaks, of the $M^L$ curve, correspond to the change from $k_{max}=1$ to $k_{max}=0$ for the \textit{up} and \textit{down} quark, respectively. As we have seen, for very low temperatures the value of $\mu$ at coexistence decreases with $B$ (see Fig 1) so that, generally,  $k_{max}$ and $M$ must vary and when $k_{max}$ decreases, $M$ increases. It then  follows, from Eq. (\ref{eq_rho_t0}), that $\rho_B$ must decrease. When $k_{max}=0$ for both quark flavors there are no further changes in the upper Landau level and the low temperature oscillations stop at $eB \gtrsim 9.5 \,  m_\pi^2$. The magnetic field seems to cause a big change on the shape of the coexistence region so that the same high density, $\rho_B^H$,  may coexist with distinct low densities, $\rho_B^L$,  for two different values of the temperature. For example, $\rho_B^H \approx 2.5 \, \rho_0$ coexists with $\rho_B^L \approx 0 $ at $T \approx 20 {\rm MeV}$ and with $\rho_B^L \approx 1.8 \, \rho_0$ at $T \approx 70 {\rm MeV}$. This pattern is also observed, although in a milder way, at the high value $eB=15 \, m_\pi^2$.  Our results also suggest that the highest density is achieved at temperatures close to $T_c$, in opposition to the $B=0$ case where this happens at $T=0$. Another interesting quantity to be investigated in connection with the density oscillation is the latent heat, $\Delta {\cal E}$, which at $T=0$ is simply given by $\Delta {\cal E}= \mu (\rho^H - \rho^L)$ since the the two coexisting densities occur at the same pressure and chemical potential. Figure 3 (right panel) shows that the latent heat value also oscillates around its $B=0$ value for $eB \lesssim 9.5 \, m_\pi^2$.
Figure 4 presents the more intuitive $P-T$ phase diagram  showing the region of broken symmetry, which corresponds to the ``gas'' phase, and the region of (partially) restored chiral symmetry, which corresponds to the ``liquid'' phase in analogy with a liquid-gas transition. The figure shows that the transition from the ``gas'' (CSB)  phase to the ``liquid'' (CSR) phase occurs at lower pressures when $B$ increases, as expected. 
Let us now look at some other thermodynamical quantities, such as  interaction measure (or trace anomaly), which is defined by
\begin{equation}
 \Delta = \frac {({\cal E} - 3P)}{T^4} \,\,.
\end{equation}
\begin{figure}[tbh]
\vspace{0.5cm} 
\epsfig{figure=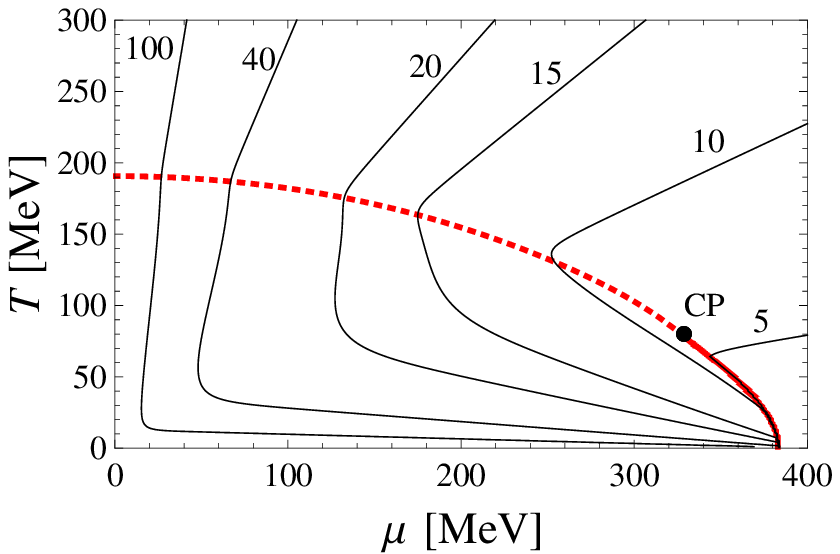,angle=0,width=7cm}
\epsfig{figure=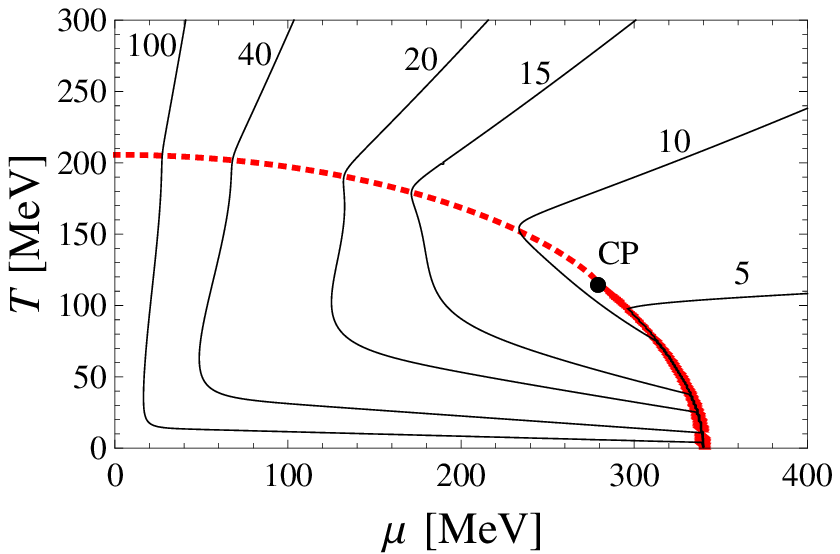,angle=0,width=7cm}
\caption{ The entropy per baryon number, $S/A$, for the NJL model at $B=0$ (left panel) and $eB=15 \,m_\pi^2$ (right panel). The thin continuous  curves correspond to $S/A=5,10,15,20,40,100$. The thick continuous lines represent first order phase transitions while the dashed lines represent crossovers. The focusing effect is not observed.}
\label{fig7}
\end{figure}
The effect of $B$ over this quantity is shown in Fig. 5 for $\mu=0$ (crossover)  and the critical chemical potential value at $B=0$, $\mu_c=328 \, {\rm MeV}$ (first order transition). The presence of a magnetic field enhances sharper transitions. Next, in order to verify how the magnetic field affects the critical region of the phase diagram, let us consider the quark number susceptibility, which is defined as
\begin{equation}
 \chi_q = \frac {d \rho}{d \mu} \,\,.
\end{equation}
Figure 6 shows this quantity, in units of $\Lambda^2$, for $B=0$ and $eB= 15 \, m_\pi^2$. In both cases one observes that the critical region set by the contour of $\chi_q$ has an elongated shape due to the enhancement of this quantity along the first order transition line. As discussed by Schaefer and Wambach \cite{wambach}, at $B=0$,  the reason for this behavior is that along the path asymptotically parallel to the first order line the quark number  susceptibility scales with the exponent $\gamma_q$ which, within the MFA, is expected to be $\gamma_q=1$ while for any other path the divergence scales with $\epsilon$, which in the MFA takes the value $\epsilon = 2/3 < \gamma_q$. Although we do not perform the explicit evaluation of the associated critical exponents here we can, nevertheless, take this explanation as being valid also at $B \ne 0$ since the shape of the critical region remains elongated along the first order transition line when a strong magnetic field is present as shown by our results. 
Finally, another interesting physical quantity, that is  easily obtained from the EoS, is the entropy per baryon  number
\begin{equation}
 \frac{S}{A} =  3 \frac{({\cal E} + P - \mu \rho)}{T\rho} \,\,.
\end{equation}
At $B=0$, the quantity $S/A$ was considered in Ref. \cite {scavenius} in order to check an eventual convergence of the adiabats toward the critical point as claimed in Ref. \cite{stephanov}. However, the authors of Ref. \cite {scavenius} did not observe such an effect and the result was explained by recalling that since there is no  change in the degrees of freedom of the two phases one should not expect the focusing effect to arise within the LSM and the NJL type of models. In this case the adiabats show the typical behavior of an ordinary liquid-gas phase transition represented, respectively, by the chirally symmetric and broken phases \cite{mishustin}.  Our results, displayed in Fig. 7,  show that the magnetic field does not produce any noticeable effect on the behavior of the adiabats which present a similar pattern at vanishing and high $B$.

\section{Thermodynamics of the  LSM under a Magnetic Field}

In this section,  we follow  Ref. \cite {andkhan} and consider the LSM in the approximation where the mesonic sector contributes only at the classical (tree) level while one loop quantum corrections, including vacuum contributions, are furnished by the quark sector alone.  We will perform our analysis considering two sets of parameter values, which produce a high and a low value for $m_\sigma$, in order to address the efficiency of the MFA in dealing with this particular model in the presence of magnetic fields.

\subsection{The Model}

The  lagrangian density of the LSM with quarks reads
\beq \label{lagsl} \mathcal{L}_{\rm LSM} = \bar{\psi}[i{\partial \hbox{$\!\!\!/$}} - g(\sigma+i\gamma_{5}\vec{\tau}\cdot\vec{\pi}) ]\psi +  \frac{1}{2}(\partial_{\mu}\sigma\partial^{\mu}\sigma + \partial_{\mu}\vec{\pi}\cdot\partial^{\mu}\vec{\pi}) - U(\sigma,\vec{\pi}) \,, \eeq
where $\psi$ is the flavor isodoublet spinor representing the quarks, and 
\begin{equation}
U(\sigma,\vec{\pi}) = \frac{\lambda}{4}(\sigma^2 + \vec{\pi}^2 - v^2)^2 - h\sigma \; ,
\label{lsm2}
\end{equation}
is the classical potential energy density. 
In the chiral limit (obtained by setting $h=0$ in the previous equation) the 
chiral symmetry $SU(2)_V\times SU(2)_A$ is spontaneously broken at the classical level, and the pion is the associated
massless Goldston boson.   
Here, we are interested in the  $h \neq 0$ case which  implies that chiral symmetry is explicitely broken giving  the pion a finite mass at $T=0$ and $\mu=0$. 
The parameters are usually chosen so that chiral symmetry is spontaneously
broken in the vacuum and the expectation values of the meson fields
are $\langle \sigma \rangle=f_\pi$ and $\langle {\vec  \pi} \rangle=0$  where
$f_\pi=93 \, {\rm MeV}$ is the pion decay constant.  When pure vacuum contributions are considered within the $\overline {\rm MS}$ 
renormalization scheme, as we do here, an arbitrary mass scale $(\Lambda_{\rm {\overline {MS}}})$ also appears in the final results. 
As discussed in the introduction, for our purposes, it  will be necessary to consider two different sets of parameters in order to get a high and a low value for $m_\sigma$. The first set is given by  $\Lambda_{\rm {\overline {MS}}}=16.48 \, {\rm MeV}$ which, together with  $v=64.29 \, {\rm MeV}$, $\lambda = 46.06$, and $h =1.77\times10^6\, {\rm MeV}^3$ ,  yields  $m_\pi=138 \, {\rm MeV}$ and $m_\sigma=600 \,
{\rm MeV}$. The second set is just like the first except for $\lambda$ and $v$ which are set to $\lambda = 36.96$ and $v=54.96 \, {\rm MeV}$ yielding $m_\sigma=450 \, {\rm MeV}$. In both cases the sigma meson mass falls within the range of the broad resonance $m_\sigma=400-800 \, {\rm MeV}$ \cite {newandersen}.

\begin{figure}[tbh]
\vspace{0.5cm} 
\epsfig{figure=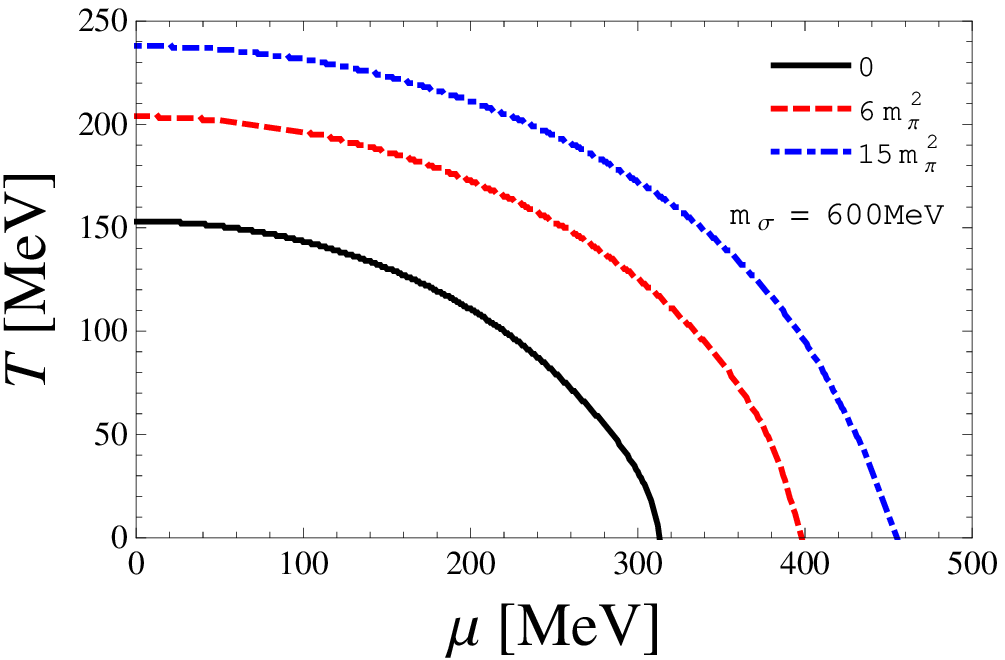,angle=0,width=7cm}
\epsfig{figure=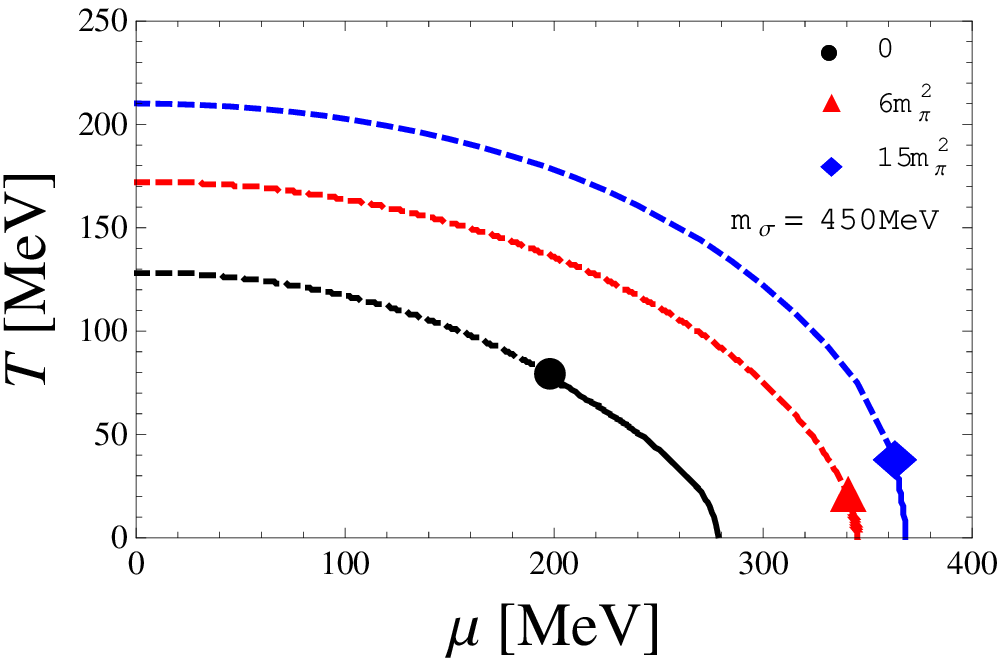,angle=0,width=7cm}
\caption{ The LSM phase diagram in the $T-\mu$ plane for  $B=0$, $eB=6\, m_\pi^2$ and $eB=15\, m_\pi^2$. The left panel, in which only crossover occurs, refers to $m_\sigma=600\, {\rm MeV}$  and the right panel to $m_\sigma=450\, {\rm MeV}$. In the right panel, the crossover is represented by dashed lines, the first order phase transitions by  continuous lines and the CP, by  solid symbols which are located at ($T_c \simeq 79.32 \, {\rm MeV}, \mu_c \simeq 198.52 \, {\rm MeV}$) for $B=0$, ($T_c \simeq 12.3 \, {\rm MeV}, \mu_c \simeq 343.4 \, {\rm MeV}$) for $eB=6 \, m_\pi^2$, and ($T_c \simeq 34.2 \, {\rm MeV}, \mu_c \simeq 365.1 \, {\rm MeV}$) for $eB=15 \, m_\pi^2$. In both cases the region of  broken  symmetry expands with increasing $B$.}
\label{fig8}
\end{figure}
\begin{figure}[tbh]
\vspace{0.5cm} 
\epsfig{figure=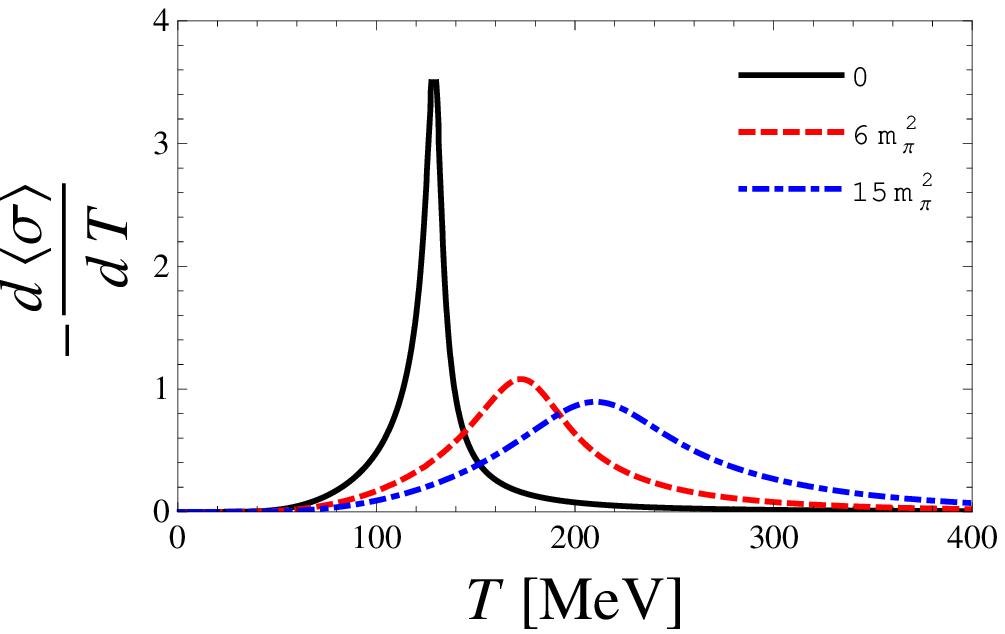,angle=0,width=7cm}
\epsfig{figure=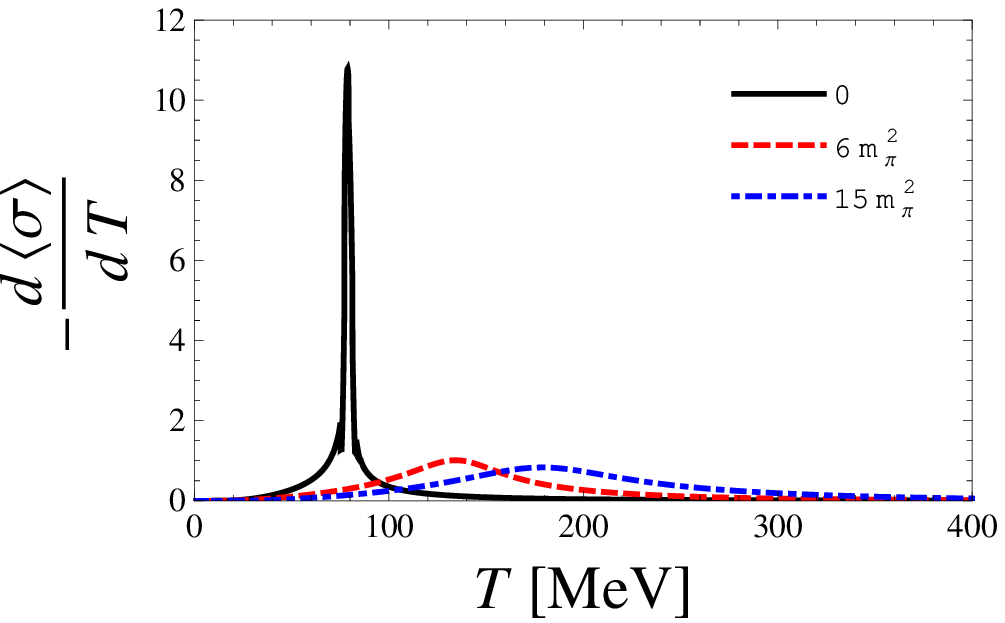,angle=0,width=7cm}
\caption{ The quantity  $-d\langle \sigma \rangle/dT$, where $\langle \sigma \rangle$ is the LSM order parameter, as a function of $T$ for $m_\sigma=450\, {\rm MeV}$. The left panel corresponds to $\mu=0$ and the right  panel to $\mu_c(B=0)=198.52\, {\rm MeV}$.}
\label{fig9}
\end{figure}

\subsection{The LSM free energy in the presence of a magnetic field}

Treating the bosonic degrees of freedom at the tree level implies that the loop contributions to the free energy come entirely  from the fermionic sector. This contribution is represented by an integral similar to the one which appears in Eq. (\ref{free}) so that the free energy  is given by 

\begin{equation}
\mathcal{F}^{\rm LSM}= U(\sigma,\vec{\pi})+\mathcal{F}_{\rm vac}^{\rm LSM}+\mathcal{F}_{\rm mag}^{\rm LSM}+
\mathcal{F}_{\rm med}^{\rm LSM} \,\, ,
\label{freelsm}
\end{equation}
where, the magnetic and the in-medium terms bear the same form as their NJL counterparts with the obvious replacement $M^2 \to m_q^2= g^2 (\sigma^2 + {\vec \pi}^2)$. On the other hand, the  pure vacuum contribution, whose general from is given by Eq. (\ref {vacnjl}), is treated in different fashion within both models and in many LSM applications it is discarded on the grounds that it does not give significant contributions at high temperature and/or chemical potential values. However, a considerable number of studies have pointed out to its importance in the characterization of the phase transition \cite {eduardo,friman,akk,andkhan}. Here, we follow Ref. \cite {andkhan} by including the pure fermionic vacuum contribution in the free energy. Since this model is renormalizable the usual procedure is to regularize divergent integrals  using  dimensional regularization and then to subtract the ultra violet divergences in the $\overline {\rm MS}$ renormalization scheme. This procedure gives  the following finite result
\begin{equation}
\mathcal{F}_{\rm vac}^{\rm LSM}=  \frac{ N_c N_f m_q^4}{(8 \pi)^2} \left [\frac{3}{2}- \ln \left ( \frac{m_ q^2}{ \Lambda_{\overline {\rm MS}}^2} \right )  \right ] \, .
\end{equation}
\begin{figure}[tbh]
\vspace{0.5cm} 
\epsfig{figure=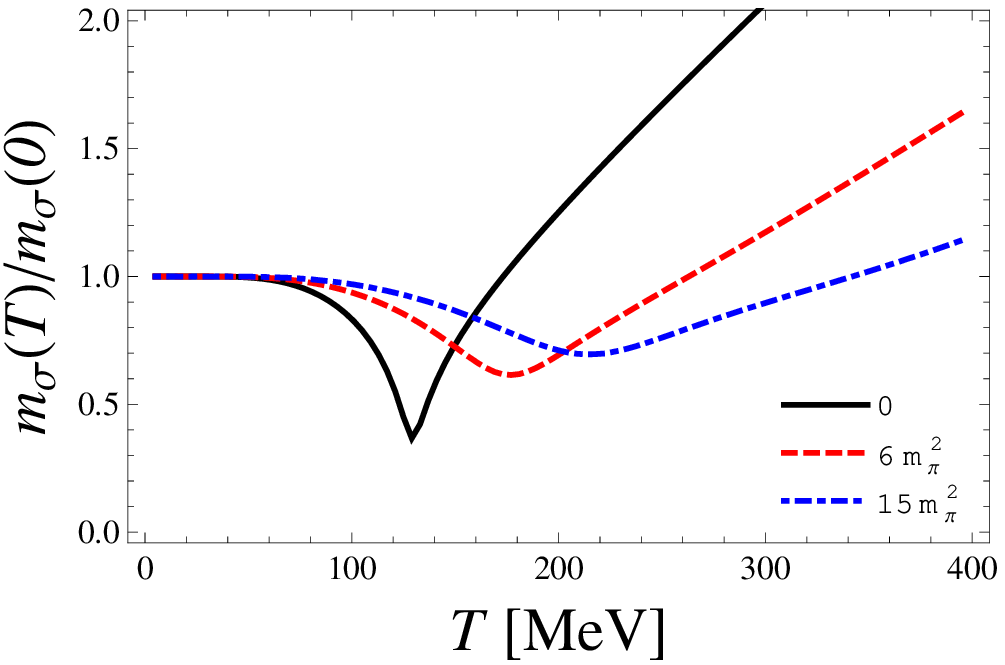,angle=0,width=7cm}
\epsfig{figure=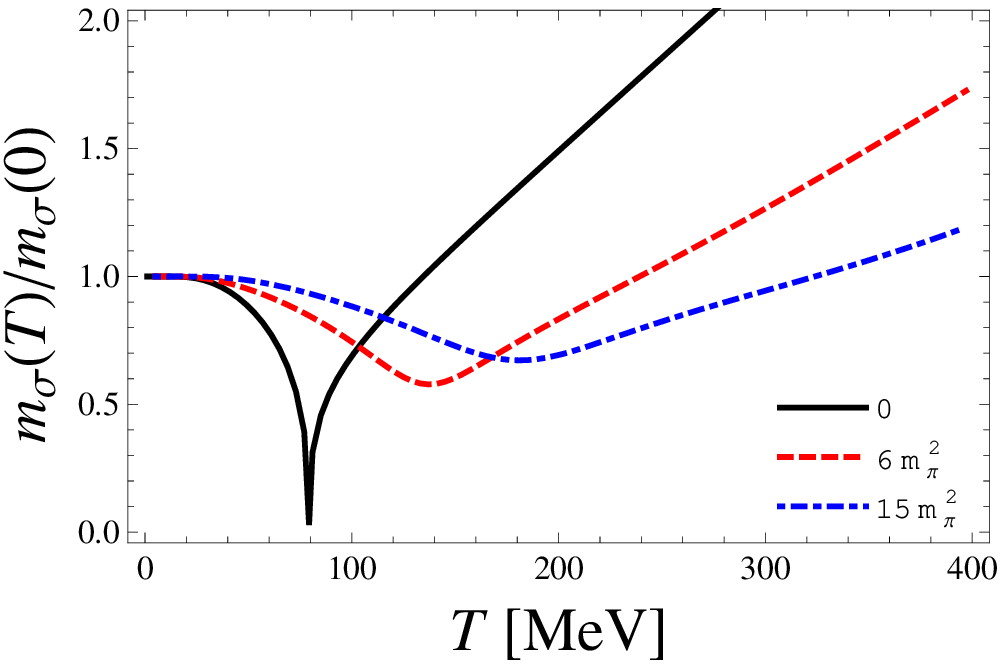,angle=0,width=7cm}
\caption{ The sigma meson mass, $m_\sigma$, whose inverse represents the correlation length, as a function of $T$ for $\mu=0$ (left panel) and $\mu_c(B=0)=198.52\,{\rm MeV}$ (right panel). For comparison reasons the mass has been normalized by its value at $T=0$ ($m_\sigma=450\, {\rm MeV}$). The curve correspondig to $B=0$ in the right panel shows that $m_\sigma \to 0$ when $\mu \to \mu_c(B=0)$, as expected. }
\label{fig10}
\end{figure}

\begin{figure}[tbh]
\vspace{0.5cm} 
\epsfig{figure=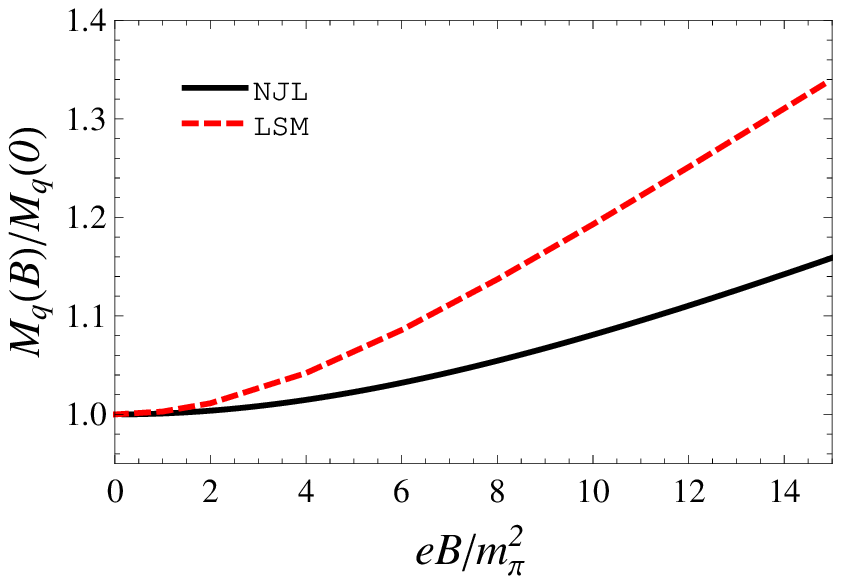,angle=0,width=7cm}
\epsfig{figure=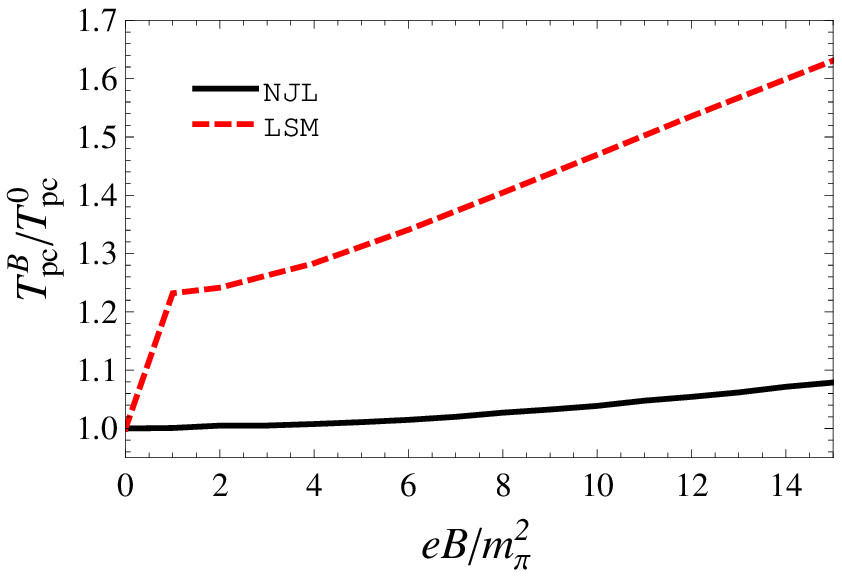,angle=0,width=7cm}
\caption{Left panel: Manifestation of magnetic catalysis, at $T=0$, for the NJL model (continuous line) and for the LSM (dashed line) with $m_\sigma=450\, {\rm MeV}$. The effective quark mass, $M_q(B)$, which represents the effective quark mass for each model, has been normalized by its $B=0$ value to allow for comparisons. Right panel: The pseudocritical temperature, as a function of $B$, for each model. This quantity has also been normalized by its value at $B=0$ for comparison reasons and the LSM is taken with $m_\sigma=450\, {\rm MeV}$. }
\label{fig11}
\end{figure}
The expectation values, $\langle \sigma \rangle $ and $\langle \vec \pi \rangle$ fields are obtained by extremizing the free energy 
\begin{equation}
 \frac{\partial \mathcal{F}^{\rm LSM}}{\partial \sigma}=0 \,\,\,,\,\,\,\frac{\partial \mathcal{F}^{\rm LSM}}{\partial \pi_i}=0 \,\,\,\,,
 \end{equation}
while the mesonic masses are determined by the curvature of the free energy at the global minimum
\begin{equation}
m^2_\sigma= \frac{\partial^2 \mathcal{F}^{\rm LSM}}{\partial \sigma^2} \,\,\,,\,\,\,m^2_{\pi_i} =\frac{\partial^2 \mathcal{F}^{\rm LSM}}{\partial \pi_i^2} \,\,\,\,,
 \end{equation}
evaluated at  $\sigma = \langle \sigma \rangle $ and $\pi = \langle \vec \pi \rangle$. Here, we follow the usual procedure  \cite {kapusta,fkp,scavenius}  by setting $\langle \vec \pi \rangle =0 $ so that $m_q = g \langle \sigma \rangle $.

\subsection{Numerical Results for the LSM}
Let us start by mapping the $T-\mu$ phase diagram, at the physical point, for three relevant values of the magnetic field ($eB=0, 6 m_\pi^2, 15 m_\pi^2$). Let us start with the parameter set which gives $m_\sigma=600 \, {\rm MeV}$ whose result is shown in the left panel of Fig. 8 which indicates that the high-$T$/low-$\mu$ behavior is similar to the one found in the  NJL  model. Namely, a crossover happens at $T_{\rm pc}$ values which increase with the magnetic field.  Note that, qualitatively, this situation  does not change even by going beyond the MFA or employing other parametrizations \cite {skokov, newandersen}. Here, the results given by the parametrization leading to $m_\sigma=600 \, {\rm MeV}$  start to depart from the ones observed in the NJL case   at intermediate values of  $T$ and $\mu$  since the crossover  does not change into  the expected first order transition with its associated CP. Then, the whole $T-\mu$ plane is dominated by the crossover exactly as the authors of Ref. \cite {andkhan} have observed using $m_\sigma=800 \, {\rm MeV}$. The situation changes drastically if one uses the parameter set which yields $m_\sigma=450 \, {\rm MeV}$ as the right panel of Fig. 8 suggests. Indeed, with this set one observes the crossover/CP/first order transition pattern. However, when the  magnetic field increases one does not observe the characteristic effects observed earlier in the NJL model as the decrease of the coexistence chemical potential at low $T$ (IMC), 
the increase of $T_c$ and the decrease of $\mu_c$ for example. 
We can further investigate the strength of the transition by analyzing how the order parameter varies with the temperature as shown in Fig. 9 for $\mu=0$ and $\mu_c(B=0)=198.52\,{\rm MeV}$. In both cases one sees that the strength, characterized by the peak of $-d\langle \sigma \rangle/dT$ decreases with increasing magnetic fields and  the figure also suggests that, when $B \ne 0$ and $\mu \ne 0$,  the crossover becomes smoother as $B$ increases. Based on the findings of Ref. \cite {nois}, which show that a magnetic field makes the transition sharper, one would expect the opposite behavior. Since the correlation length is governed by $1/m_\sigma$ \cite {correlation}   the sigma meson mass, as a function of $B,T$ and $\mu$ is also an interesting quantity to be analyzed and with this aim we present Fig. 10.  In Ref. \cite {skokov}, which considers the PQM at $\mu=0$,  it has been demonstrated that the value of $m_\sigma$ at $T_{\rm pc}$ is almost independent of $B$ (both, in the MFA and FRG approaches). Contrary to that, within our approximation, we observe that  the difference $m_\sigma(T_{\rm pc}) -m_\sigma(0)$ decreases with increasing $B$ which is also in opposition to the expectations furnished by the results of Ref. \cite {nois}. Within the NJL model the behavior of $m_\sigma(T,\mu,B)$ has also been recently addressed \cite {mesons} furnishing results which predict that $m_\sigma(T_{\rm pc}) -m_\sigma(0)$ increases with increasing $B$ and also that the transition appears to be sharper as $\mu$ increases. These comparisons suggest that the naive application of the MFA to the LSM at finite $T,\mu$ and $B$ indeed misses some important physics, especially at finite densities. This conclusion is not only supported by the NJL-MFA results of Refs. \cite {inagaki,andreas,imc,nois,mesons} but also by the LSM-FRG results obtained by Andersen and Tranberg \cite {newandersen}.

\subsection{Comparing Magnetic Catalysis in the LSM and the NJL model at $\mu=0$}

In the previous subsection we have pointed out that the MFA does not seem to furnish reliable results for the LSM. However, this issue seems to be less severe at $\mu=0$ where at least the predicted type of transition is in agreement with other model applications as well as with lattice simulations. We can then address the question of how magnetic catalysis \cite {MC} (see Ref. \cite{igor} for an updated review) affects each model by comparing the dimensionless quantities $T_{\rm pc}^B /T_{\rm pc}^0$ and $M_q(B)/M_q(0)$ where, now, $M_q$  is representing the effective quark mass in each case, that is, $M$ in the NJL model and $m_q$ in the LSM. The result, shown in Fig. \ref {fig11}, suggests that within our approximations the effective quark mass of the LSM is more sensitive to the presence of a magnetic field. At $eB=15\,m_\pi^2$ the LSM quark mass has increased by about $20 \%$ while the NJL quark mass has increased by about $15 \%$. The increase of $T_{\rm pc}$ is even more dramatic within the LSM and at   $eB=15\,m_\pi^2$ it increases by about $50 \%$ in relation to the $B=0$ value while in the NJL the increase  is about $9 \%$. At low values of $B$ the LSM pseudocritical temperature suffers a sudden increase which again may be an indication of the failure of the approximations we have adopted for this particular model.

\section{Conclusions}

We have considered both, the LSM and the NJL model, in the description of hot and dense two flavor quark matter subject to strong magnetic fields, such as the ones expected to  be created in non central heavy ion collisions.  Here, the simplest version of these models, which do not consider the Polyakov loop, have been treated within the MFA. In this approximation,  one loop fermionic contributions have been considered,  allowing for the determination of the phase diagram associated with chiral symmetry breaking/restoration. So far, most applications  \cite {inagaki, andkhan, nois, newandersen} at finite $T,\mu$ and $B$  have mainly focused in  the determination of the $T-\mu$ phase diagram  without analyzing the effects of the magnetic fields over  important physical quantities such as the coexistence boundaries,  the quark number susceptibilities, the interaction measure and the mesonic masses among others. Therefore, our aim was to perform this analysis in order to understand the effects of $B$ in more detail, especially at low temperatures, which is the less explored region. We have started by considering the NJL model, with physical quark masses,  obtaining a $T-\mu$ phase diagram which is in qualitative agreement with the one obtained in Ref. \cite {nois} for three quark flavors and also with Ref. \cite{mesons} for two quark flavors.  One of the most important features of this diagram is related to the position of the critical end point that appears to be located at higher temperatures and smaller chemical potentials when $B$ increases. One also observes that the first order segment of the  transition line increases with $B$ while the coexistence chemical potential value decreases at low temperatures. These observations suggest that the magnetic field has a direct effect in the physics associated with the first order transition and may influence the size and shape of the coexistence region for example. In order to check that we have mapped the $T-\mu$ plane into the $T-\rho_B$ plane showing that for $eB \lesssim 9.5 \, m_\pi^2$ the high density branch of the coexistence phase diagram oscillates around its $B=0$ value as a consequence of filling the Landau levels which influences the values of quantities such as the latent heat as we have shown. This finding may also have consequences regarding, e.g., the physics of phase conversion whose dynamics requires the knowledge of the EoS {\it inside} the coexistence region. For example, at a given temperature, the surface tension between the two coexisting bulk phases ($\rho^L$ and $\rho^H$) depends on the value of their difference \cite {jorgen,surten} which will, possibly, be affected by the oscillations suffered by the coexistence boundary due to the presence of a magnetic field. This feature of the phase diagram  deserves further investigation since it may be important in the description of the EoS inside the boundaries.
Then, using the quark number susceptibility, we have mapped  the critical region  observing that its shape, which is elongated along the first order transition line, is not affected by the presence of a magnetic field. We have also investigated the behavior of the adiabats in the presence of $B$ observing the same pattern which was observed at $B=0$ \cite {scavenius}. The two latter results indicate that general model characteristics, associated with the shape of the critical region and the isentropic trajectories, are unaffected by the presence of $B$. For the NJL model we conclude that  magnetic fields appear to cause a significant effect at the intermediate and low temperature parts of the phase diagram causing the crossing of the transition lines representing different values of $B$. The presence of magnetic fields also promotes a change on the size and location of the first order line. 

Our results for the LSM indicate that the MFA treatment as performed here does not seem to be appropriate, especially at higher values of the chemical potential where, as discussed, the magnetic field has a strong influence. In the light of the results furnished by the more powerful FRG  \cite{skokov,newandersen} and also those obtained with the NJL model within the MFA \cite{inagaki, andreas,imc,nois,mesons} we have observed a significant disagreement, which include the absence of IMC in the LSM phase diagram, in the low temperature region.

\section*{Acknowledgments}

MBP was partially supported by CNPq. GNF and AFG thank Capes and CNPq for MSc scholarships. We would like to thank Jens Andersen and Sandeep Chatterjee for discussions related to the LSM.

\end{document}